\documentstyle [prl,aps]{revtex}
\newcommand{\be}{\begin{equation}}
\newcommand{\ee}{\end{equation}}
\newcommand{\bea}{\begin{eqnarray}}
\newcommand{\eea}{\end{eqnarray}}

\newcommand{\ri}{\mbox{i}}
\newcommand{\re}{\mbox{e}}

\begin {document}
\bibliographystyle {plain}
\begin{titlepage}
\begin{flushright}
\end{flushright}
\vspace{0.5cm}
\begin{center}
{\Large {\bf Phase transitions in the one-dimensional
spin-S $J_1-J_2$ XY model}}\\
\vspace{1.8cm}
\vspace{0.5cm}
{P. Lecheminant$~^{1}$, T. Jolicoeur$~^{2}$, P. Azaria$~^{3}$}\\ 
\vspace{0.5cm}
{\em$^{1}$ 
Laboratoire de Physique Th\'eorique et Mod\'elisation,\\
Universit\'e de Cergy Pontoise, 5 Mail Gay-Lussac,
Neuville sur Oise, 95301 Cergy
Pontoise Cedex, France}\\                               
{\em$^{2}$ Laboratoire de Physique de la Mati{\`e}re Condens{\'e}e, 
Ecole Normale Sup{\'e}rieure, 24 rue Lhomond, 75005 Paris, France} \\ 
{\em$^{3}$
Laboratoire de Physique Th\'eorique des Liquides,\\
Universit\'e Pierre et Marie Curie, 4 Place Jussieu, 75252 Paris, France}
\vspace{2cm}
\begin{abstract}  
The one-dimensional spin-S $J_1-J_2$ XY model is  
studied within the bosonization approach.
Around the two limits ($J_2/J_1 \ll 1,J_2/J_1 \gg 1$)
where a field theoretical analysis can be derived,
we discuss the phases
as well as the different phase transitions that occur
in the model.
In particular, it is found that
the chiral critical spin nematic phase, 
first discovered by Nersesyan et al. (Phys. Rev. Lett. {\bf 81},
910 (1998)) for S=1/2,
exists in the general spin-S case. 
The nature of the effective field theory that 
describes the transition between this chiral
critical phase and a chiral gapped phase is 
also determined. 
\end{abstract}
\end{center}
{\rm PACS No: 75.10.Jm} 
\end{titlepage}
\sloppy
\par                                 
\section{Introduction}

The interplay between frustration and quantum fluctuations
in low-dimensional spin systems has attracted much
interest. 
One of the main reasons for this attention is that
frustration is expected to lead to 
new exotic phases as well as unconventional spin excitations.
In the one dimensional case, powerful non-perturbative methods are
available and the key features of frustration 
can then be analysed in depth. 
In this respect, the 
phase diagram of the one dimensional S=1/2 $J_1-J_2$ XY model 
has been studied extensively over the years. 
The lattice Hamiltonian of this model is defined by:
\be
{\cal H} =  J_1 \sum_n\left(S^x_n S^x_{n+1} + S^y_n S^y_{n+1}\right)
+ J_2 \sum_n\left(S^x_n S^x_{n+2} + S^y_n S^y_{n+2}\right), 
\label{hamxyfrus}
\ee
where $S_n^{\pm} = S_n^x \pm i S_n^y$ is 
a spin-1/2 operator at site n and $J_2$ is a competitive
antiferromagnetic interaction ($J_1,J_2 > 0$) which introduces 
frustration in the model.
The phase diagram of the Hamiltonian (\ref{hamxyfrus}) 
is expected to be rich.
For a small value of $J_2$, one has a spin fluid phase (XY phase)
characterized by gapless excitations with central charge $c=1$ 
whereas for $J_2/J_1 \simeq 0.32$\cite{nomura} a phase transition of 
Kosterlitz-Thouless (KT) type
occurs and the model enters a massive region 
with a two-fold degenerate ground state (dimerized phase)\cite{haldane}. 
Interestingly enough,
it was recently predicted by 
Nersesyan et al.\cite{nersesyan}, within 
the bosonization approach\cite{book2}, that in the large $J_2$ limit,
where the model can be viewed as a two-leg XY zigzag ladder, 
a critical spin nematic phase 
with chiral long-range order 
($\langle ({\vec S}_n \wedge
{\vec S}_{n+1})_z \rangle \ne 0$)  should emerge. 
This unconventional 
phase with unbroken time reversal symmetry is characterized by  
nonzero local spin currents polarized along the anisotropy $z$-axis.
The transverse spin-spin correlation functions are incommensurate
and fall off with the distance as a power law with the exponent $1/4$. 
However, this phase has not been reported in the numerical calculations
of Refs. \cite{kawa,kawa1}. 
In contrast, only the  previous
well known phases (the spin fluid and dimerized phases)
have been found.
A numerical analysis of the model 
using an exact diagonalization method 
with twisted boundary conditions has pointed out the presence of 
incommensuration in the large $J_2$ limit\cite{aligia} but
within this approach it has not been possible  
to conclude on the criticality or not of this phase.
Very recently, Nishiyama\cite{nishi} has investigated the
existence of chiral order of the Josephson-junction 
ladder with half a flux quantum per plaquette by
means of the exact diagonalization method.
He was able to show that the critical
phase predicted in Ref. \cite{nersesyan} does 
exist in the range of the parameters of the model (\ref{hamxyfrus})
in constrast to the numerical findings of Ref. \cite{kawa}.

The situation is less controversial 
in the S=1 case 
and the corresponding phase diagram 
has been determined by a DMRG study \cite{kawa,kawa1}. 
The model  
with $J_2=0$ is a critical 
spin fluid (the so-called XY1 phase\cite{schulz})
and as soon as the next-nearest-neighbor interaction is switched on 
the Haldane phase is stabilized (KT transition).
As noted in Ref. \cite{kawa},
this fact seems to be in contradiction with the bosonization result 
obtained in the small $J_2$ limit\cite{shimaoka} which suggests
that the XY1 phase extends to finite $J_2/J_1$.
Increasing on the value of $J_2/J_1$, the authors of Refs. \cite{kawa,kawa1}
have reported the occurence of two successive transitions:
A first one at $(J_2/J_1)_{c1} \simeq 0.473$ (Ising transition) from 
the Haldane phase to a gapped phase 
with chiral long-range order (chiral gapped phase)
and a second transition at  $(J_2/J_1)_{c2} \simeq 0.49$ (presumably
a KT transition) with a chiral critical phase which corresponds
to the spin nematic phase discussed by Nersesyan et al.\cite{nersesyan}
in the context of the two-leg S=1/2 XY zigzag ladder.
In this latter phase, the spin-spin correlation are incommensurate
and decay with a power law with an exponent 
approximatively equals to $0.15$\cite{kawa1}.
Recently, the phase diagram 
of the model (\ref{hamxyfrus}) in the general spin-S case has been 
further discussed
in Ref. \cite{kol} by means of a large-S approach.
It has been found that the existence of the 
gapless and gapped chiral phases
is not specific of S=1 but is rather a generic large-S feature.
The predicted phase diagram has four different phases.
First of all, one has
the XY1 critical spin fluid phase and
for a finite value of $J_2$ (KT transition) 
one enters the Haldane phase.  
As $J_2/J_1$ is further increased, there are 
the two successive phase transitions previously
discussed: An Ising-type transition separating the Haldane
phase from the chiral gapped phase, 
followed by a KT transition corresponding to 
the transition between the chiral gapless phase and 
the chiral gapped phase.
However, this large-S approach does not distinguish between integer
and half-integer spins and also it does not take into account the 
possibility of a spontaneously dimerized phase. 
In addition,
for the very special S=1 case,
the numerical calculations of Refs. \cite{kawa,kawa1}
predict rather that the critical
phase XY1 does not extend for a finite value of $J_2$.

In this paper, we shall investigate the phase diagram of
the one-dimensional spin-S $J_1-J_2$ XY model 
within the bosonization approach.
Using the Abelian bosonization of a general spin-S operator introduced
by Schulz\cite{schulz}, the low-energy physics of 
the Hamiltonian (\ref{hamxyfrus}) can be studied 
in two different limits:
in the weak coupling limit when $J_2/J_1 \ll 1$ and in the ladder
limit $J_2/J_1 \gg 1$.                        
This enables us to determine the nature of the 
phase transitions that occur in the model 
from a field theoretical point of view together with a comparison with
the numerical results\cite{kawa,kawa1} as well as the 
large-S preditions\cite{kol}. The remainder of the paper 
is organized as follows: the weak coupling analysis ($J_2/J_1 \ll 1$)
is given in Section II whereas the zigzag limit of the model
($J_2/J_1 \gg 1$) is performed in Section III.
Section IV presents our concluding remarks and finally
the conventions and some technical details used in this 
work are described in the two 
appendices.

\section{Weak coupling limit} 

In this section, we shall investigate the low energy
physics of the one-dimensional spin-S $J_1-J_2$ XY model
in the limit $J_2/J_1 \ll 1$ within the bosonization approach.
This enables us to study 
the stability of the critical XY1 phase 
(for $J_2=0$) upon switching on a small next-nearest-neighbor
interaction.
The Hamiltonian (\ref{hamxyfrus}) in the S=1 case
can be tackled within the bosonization 
framework by 
representing spin-1 operators as a sum of two spin-1/2 operators:
$S^{\pm}_n = s^{\pm}_{1,n} + s^{\pm}_{2,n}$.
It has been argued\cite{luther,nijs,solyom}
that the additional local singlets introduced will lead to extra levels
with higher energy than the triplet states so that the ground state
and the low-lying excitations are correctly captured. 
In particular,  with this representation, Timonen and Luther\cite{timonen} 
and Schulz\cite{schulz} predicted the correct phase diagram 
of the one-dimensional anisotropic antiferromagnetic spin-1
Heisenberg model with single-ion anisotropy.
Moreover, the effect of weak randomness on this latter 
model has also been analysed 
within this bosonization approach\cite{brunel}.
This procedure was further generalized by 
Schulz\cite{schulz} by representing a general 
spin-S operator $S^{\pm}_n$ as the sum of 2S spin-1/2
operators $s^{\pm}_{a,n}, a=1,..,2S$: 
\be 
S^{\pm}_n  = \sum_{a=1}^{2S} s^{\pm}_{a,n},
\label{decompospin}
\ee
from which the low-energy physics of the one-dimensional spin-S 
Heisenberg model can be captured and in particular
the difference between half-integer 
and integer spins\cite{haldanebis}.

Let us first review the continuum description of the spin-S XY 
chain first obtained by Schulz\cite{schulz}.  
Using the decomposition (\ref{decompospin}),
the Hamiltonian (\ref{hamxyfrus}) for $J_2=0$ writes:
\be
{\cal H}_{XY1} =  \frac{J_1}{2} \sum_{a=1}^{2S}
\sum_n\left(s^{\dagger}_{a,n} s^{-}_{a, n+1} + H. c.\right)
+ \frac{J_1}{2} \sum_{a \ne b} \sum_n\left(s^{\dagger}_{a,n} s^{-}_{b,n+1}
+ H. c.\right).
\label{spinSxy}
\ee
The first term in this equation corresponds to $2S$ decoupled 
spin-1/2 XY chains. 
As recalled in the Appendix A,
the low-energy physics of the spin-1/2 XY chain 
can be extracted from the introduction of  
a single U(1) bosonic field $\varphi$
with chiral components $\varphi_{R,L}$.
As a consequence, the next step of the approach is 
to introduce
2S chiral decoupled bosonic fields $\varphi_{aR,L}, a=1,..,2S$ 
and the Hamiltonian density of the first term in Eq. (\ref{spinSxy})
in the continuum limit is given by:
\be 
{\cal H}_{XY} \simeq \frac{v_0}{2}\sum_{a=1}^{2S}\left(
\left(\partial_x \varphi_a\right)^2
+ \left(\partial_x \vartheta_a\right)^2\right), 
\ee
where $\varphi_a = \varphi_{aL} + \varphi_{aR}$, 
$\vartheta_a = \varphi_{aL} - \varphi_{aR}$ being the
dual field, and $v_0=J_1 a_0$ ($a_0$ being
the lattice spacing) is the spin velocity.
One can then derive the continuum
limit of the second term in Eq. (\ref{spinSxy})
using   
the bosonic description (\ref{contspinxy}) of the
spin-1/2 operators $s_{a}^{\pm}$ described in the Appendix A:
\bea
s^{+}_a =  \frac{\left(-1\right)^{x/a_0}}{\sqrt{2\pi a_0}}
\exp\left(\ri \sqrt{\pi}\vartheta_a\right) \nonumber \\
+ \frac{1}{\sqrt{8\pi a_0}}
\left(\exp\left(\ri 3 \sqrt{\pi} \varphi_{aL}
+ \ri \sqrt{\pi} \varphi_{aR} \right)
+\exp\left(-\ri 3 \sqrt{\pi} \varphi_{aR} - \ri \sqrt{\pi} \varphi_{aL} \right)
\right).
\label{contspinxySbis}
\eea
Notice that a priori this procedure has only a sense provided that 
the coupling
constant associated to the second piece of 
Eq. (\ref{spinSxy}) is much smaller than $J_1$
and this is clearly not the case here. However, one expects
in this problem, on
general grounds, a continuity between weak and strong coupling
limits so that it is natural to bosonize the
second term of Eq. (\ref{spinSxy}) using
the correspondence (\ref{contspinxySbis}).
The leading part of the density Hamiltonian associated
to (\ref{spinSxy}) in the continuum limit reads thus as follows: 
\bea
{\cal H}_{XY1} \simeq \frac{v_0}{2}\sum_{a=1}^{2S}\left(
\left(\partial_x \varphi_a\right)^2
+ \left(\partial_x \vartheta_a\right)^2\right)
-\frac{J_1}{2\pi a_0} \sum_{a \ne b}
\cos\left(\sqrt{\pi}\left(\vartheta_a - \vartheta_b \right) \right).
\label{hamxyS}
\eea
It is then suitable to switch to a basis to single out
the degrees of freedom that will remain critical in 
the infrared limit.
To this end,
let us introduce a diagonal bosonic field $\Phi_{+R(L)}$ and $2S - 1$
relative bosonic fields $\Phi_{mR(L)}$, $m=(1,...,2S-1)$ as follows:
\bea
\Phi_{+R(L)} &=&\frac{1}{\sqrt{2S}} \left(\varphi_{1} + ...
+ \varphi_{2S} \right)_{R(L)}
 \nonumber \\
\Phi_{mR(L)} &=&\frac{1}{\sqrt{m(m+1)}}\left( \varphi_{1} + ...+\varphi_{m}
- m \varphi_{m+1}\right)_{R(L)} .
\label{can}
\eea
The transformation (\ref{can}) is canonical and preserves the bosonic
commutation relations. This basis has been introduced in Ref. \cite{assaraf}
in the Abelian bosonization study of the one-dimensional Hubbard model
with a SU(N) symmetry.
The inverse transformation of Eq. (\ref{can}) is easily found to be:
\bea
\varphi_{1R(L)}= \frac{1}{\sqrt{2S}}\Phi_{+R(L)} + \sum_{l=1}^{2S - 1}
\frac{\Phi_{lR(L)}}{\sqrt{l(l+1)}} \nonumber \\
\varphi_{aR(L)}= \frac{1}{\sqrt{2S}}\Phi_{+R(L)}
- \sqrt{\frac{a-1}{a}} \Phi_{(a-1)R(L)}\nonumber \\
+ \sum_{l=a}^{2S - 1}
\frac{\Phi_{lsR(L)}}{\sqrt{l(l+1)}}, \;\;
a=2,...,2S - 1 \nonumber \\
\varphi_{2S R(L)}= \frac{1}{\sqrt{2S}}\Phi_{+R(L)}
- \sqrt{\frac{2S - 1}{2S}} \Phi_{(2S - 1)R(L)}.
\label{caninv}
\eea
Using Eq. (\ref{hamxyS}), one observes that all
relative dual fields $\Theta_m, m=1,..,2S - 1$ are
pinned whereas the diagonal bosonic field
$\Phi_+$ ($\Phi_+ =  \Phi_{+L} + \Phi_{+R}$) is a strongly
fluctuating field so that:                      
\be
{\cal H}_{XY1} \simeq \frac{v_0}{2}\left(
\left(\partial_x \Phi_{+}\right)^2
+ \left(\partial_x \Theta_+\right)^2\right),
\label{hamxySfin}
\ee
where $\Theta_+ =  \Phi_{+L} - \Phi_{+R}$.
From Eq. (\ref{contspinxySbis}) and by
integrating out the massive degrees of freedom, the expression
of the effective spin-S density $S^{\pm}$ in terms of the massless
bosonic field in the + sector can be deduced:
\be
S^{\pm} \sim \left(-1\right)^{x/a_0}
\exp\left(\pm \ri \sqrt{\pi/2S}\; \;\Theta_+\right).
\label{spinSdensXY}
\ee
The dual field $\Theta_+$ is thus a compactified bosonic field
with radius ${\tilde R}_S = \sqrt{2S/\pi}$.
Using the general relation between the radius ($2\pi {\tilde R}_S R_S =1$),
we deduce that the compactified radius of the bosonic field 
$\Phi_{+}$ is: $R_S =1/\sqrt{8\pi S}$. 
Futhermore, one deduces from Eq. (\ref{spinSdensXY}) that the 
transverse spin-spin correlation function has a power law
behavior with an exponent $\eta_{\perp} = 1/(4S)$. 
The value of this exponent coincides with the prediction 
of Alcaraz and Moreo\cite{alcaraz}
who have analysed the critical properties of the XXZ spin-S Heisenberg 
model by means of a combination of conformal invariance 
and exact diagonalizations techniques.
It is worth noting that the value of the exponent 
in the S=1 case
($\eta_{\perp} = 1/4$)
has been predicted by Kitazawa et al.\cite{kitazawa}
within a level spectroscopy analysis of the S=1 bond-alternating XXZ spin chain.
Finally, one 
should observe that the uniform part of the spin density is a short-ranged
piece since the massive modes that enter in this expression
have a zero vacuum expectation value and thus
give rise to an exponential decay in the uniform part
of the spin-spin correlation function. However, as shown
by Schulz\cite{schulz}, in the special half-integer case,
higher orders of perturbation theory produce
a strongly fluctuating piece in the uniform part
of the spin density (\ref{spinSdensXY}) with
scaling dimension $2S + 1/(8S)$ which is less
relevant than the alternating contribution in Eq. (\ref{spinSdensXY})
which has scaling dimension $1/(8S)$.
 
With all these results at hands, the 
stability of this critical XY1 phase with respect to a
next-nearest-neighbor exchange interaction $J_2/J_1 \ll 1$
can be analysed.
To this end, let us first rewrite the second term (${\cal H}_{XY2}$)
of the Hamiltonian (\ref{hamxyfrus})
in terms of the
2S spin-1/2 operators:
\bea
{\cal H}_{XY2} =  \frac{J_2}{2} \sum_{a=1}^{2S}
\sum_n\left(s^{\dagger}_{a,n} s^{-}_{a,n+2} + H. c.\right)
+ \frac{J_2}{2} \sum_{a\ne b}\sum_n\left(s^{\dagger}_{a,n} s^{-}_{b,n+2} +
H. c.\right).
\label{spinSxy2}
\eea
The first part of this equation corresponds to 
the sum of N decoupled next-nearest-neighbor S=1/2 XY chains.
The resulting continuum limit has been obtained by 
Haldane in the erratum of Ref. \cite{haldane}
and is reviewed for completeness in the Appendix B
(see in particular Eq. (\ref{contxy2})).
The continuum limit of the second term in Eq. (\ref{spinSxy2})
can be obtained using the bosonized
description (\ref{contspinxySbis}) of the spin density $s_a^{\pm}$.
The resulting continuum limit of the Hamiltonian (\ref{spinSxy2}) reads
thus as follows:
\bea
{\cal H}_{XY2} \simeq -\frac{4J_2 a_0}{\pi} \sum_{a=1}^{2S}
\left(\partial_x \vartheta_a\right)^2
+ \frac{J_2}{\pi a_0} \sum_{a<b} \cos\left(\sqrt{\pi}
\left(\vartheta_a - \vartheta_b\right)\right) \nonumber \\
- \frac{J_2}{\pi^2 a_0} \sum_{a=1}^{2S} \cos\left(\sqrt{16\pi}
\varphi_a\right)
+ \frac{J_2}{2\pi a_0} \sum_{a<b}
\cos\left(\sqrt{4\pi}
\left(\varphi_a + \varphi_b\right)\right)
\cos\left(\sqrt{\pi}
\left(\vartheta_a - \vartheta_b\right)\right).
\label{contspinSxy2}
\eea
Using the canonical transformation (\ref{caninv}),
one finally obtains                         
the following effective Hamiltonian:
\be
{\cal H} \simeq \frac{v}{2}\left(K \left(\partial_x \Theta_+\right)^2
+ \frac{1}{K} \left(\partial_x \Phi_+\right)^2 \right),
\label{contspinStot}
\ee
with
\bea
v &=& v_0 \sqrt{1 - \frac{8 J_2}{\pi J_1}} \nonumber \\
K &=& \sqrt{1 - \frac{8 J_2}{\pi J_1}} .
\label{luttSpara}
\eea
Of course, as usual, these latter identifications hold only in the
vinicity of the gaussian fixed point at $J_2/J_1 =0$.
However, one should carefully
look at the higher order corrections of perturbation theory
that might generate an additional operator in
the effective theory (\ref{contspinStot})
and potentially destablizes the XY1 phase.
In this respect,  integer and half-integer
spins should be treated separately.
For integer spins S, the last term in Eq. (\ref{contspinSxy2})
leads to the following contribution at the Sth order of
perturbation theory:
\be
\int \prod_{i=1}^{S} d^2 x_i  \prod_{i=1}^{S}
\cos\left(\sqrt{4\pi} \left(\varphi_{2i-1}
+ \varphi_{2i}\right)\right)\left(x_i\right)
\cos\left(\sqrt{\pi} \left(\vartheta_{2i-1}
- \vartheta_{2i}\right)\right)\left(x_i\right),
\ee
which, after integrating out the short-ranged degrees of freedom
and using the canonical transformation (\ref{caninv}),
gives rise to a fluctuating field in the $+$ channel:
$\cos(\sqrt{8\pi S} \Phi_+)$.
In the same way, for half-integer spins S, one has
the following contribution at the 2Sth order of
perturbation theory:
\be
\int \prod_{i=1}^{2S} d^2 x_i  \prod_{i=1}^{2S}
\cos\left(\sqrt{4\pi} \left(\varphi_{i}
+ \varphi_{i+1}\right)\right)\left(x_i\right)
\cos\left(\sqrt{\pi} \left(\vartheta_{i}
- \vartheta_{i+1}\right)\right)\left(x_i\right),
\ee
with the identification: $\varphi_{2S+1} = \varphi_{1}$
and $\vartheta_{2S+1} = \vartheta_{1}$.
One then obtains the following operator:
$\cos(\sqrt{32\pi S} \Phi_+)$ after averaging on the 
short-ranged degrees of freedom.
The effective field theory associated to 
the spin-S $J_1-J_2$ XY chain in the weak coupling
limit is thus:
\bea
{\cal H} \simeq
\frac{v}{2}\left(K \left(\partial_x \Theta_+\right)^2
+ \frac{1}{K} \left(\partial_x \Phi_+\right)^2 \right)
- \frac{g_{eff}}{a_0} \cos\left(\mu \sqrt{8\pi S} \Phi_+\right),
\label{effieldthS}
\eea
with $\mu = 1$ (respectively $2$) if S is integer
(respectively half-integer).
One should note that the Hamiltonian (\ref{effieldthS})
corresponds to the effective field theory of the spin-S XXZ
Heisenberg chain derived by Schulz\cite{schulz}.
In fact, the last operator in Eq. (\ref{effieldthS})
can also be justified from a symmetry analysis.
Indeed, under the one-step translation symmetry,
the bosonic field $\varphi_a$ transforms according
to (see Eq. (\ref{trans}) of the Appendix A):
$\varphi_a \rightarrow \varphi_a + \sqrt{\pi}/2 + p_a \sqrt{\pi}$,
$p_a$ being integer. From the definition (\ref{can}) of the   
diagonal bosonic field $\Phi_+$, one thus has:
\be
\Phi_+ \rightarrow \Phi_+ + \sqrt{\frac{\pi S}{2}} + p \sqrt{\frac{\pi}{2S}},
\ee
from which we conclude that the $\cos\left(\mu \sqrt{8\pi S} \Phi_+\right)$
term is the operator invariant under the translation symmetry
with the smallest scaling dimension.
 
The phase diagram of the spin-S $J_1-J_2$ XY chain in the
small $J_2/J_1$ limit can then be deduced from the structure
of the effective field theory (\ref{effieldthS}).
For a small value of $J_2/J_1$, the cosine operator in Eq. (\ref{effieldthS})
is a strongly irrelevant contribution 
and the system is critical 
with central charge
$c=1$ (Luttinger liquid): 
it is the spin fluid XY1 phase that extends to a finite value
of $J_2$.
As $J_2/J_1$ increases, 
one expects from Eqs. (\ref{luttSpara}, \ref{effieldthS})
a KT phase transition from this spin fluid phase to
a fully massive region (dimerized or Haldane phases depending
on the nature of the spin S\cite{comment}). At the transition,
the Luttinger parameter $K_c$ is equal to: $K_c = 1/(S \mu^2)$ and
a very naive estimate of the critical value of $(J_2/J_1)_c$
can then be deduced from Eq. (\ref{luttSpara}) within the bosonization
approach:
\be
\left(\frac{J_2}{J_1}\right)_c \simeq
\frac{\pi \left(S^2 \mu^4 -1\right)}{8 S^2 \mu^4}.
\label{crij2j1S}
\ee
In particular, for S=1/2,
one finds $(J_2/J_1)_c = 3\pi/32 \simeq 0.2945$
which is not too bad
in comparaison to the value obtained in the numerical simulations
of Ref. \cite{nomura}:
$(J_2/J_1)_c \simeq 0.3238$.                                              
Moreover, in the S=1 case, the XY1 phase is destabilized
upon switching on a nonzero value of $J_2$
in full agreement with the numerical findings 
of Refs. \cite{kawa,kawa1}. 
The origin of the discrepancy noted in Ref. \cite{kawa}
between the DMRG study\cite{kawa,kawa1} and the bosonization
results obtained in Ref. \cite{shimaoka} stems from the fact that
the latter authors do not look at higher orders
in perturbation theory 
as in this work.
The S=1 case 
does not correspond to the generic 
situation since we observe from
the estimate (\ref{crij2j1S}) that the size
of the XY1 phase increases as S increases in 
the half-integer and integer cases.
In this respect, one should note that the situation is in 
close parallel to the phase transition 
between the XY1 and the Haldane phases
in the integer spin-S XXZ Heisenberg chain.
In the S=1 case, the resulting phase transition occurs precisely 
at the XY1 point\cite{schulz,rommelse,kitazawa} whereas
the XY1 phase extends considerably 
as S increases\cite{schulz,alcaraz,joli2,nomura2}.

\section{The zigzag ladder limit}

We shall now study the model (\ref{hamxyfrus}) in the ladder limit
$J_1 \ll J_2$ where it can be viewed as a two-leg spin-S XY ladder
coupled in a zigzag way.
For S=1/2 Heisenberg spins, the effect of a transverse zigzag 
interchain interaction has been extensively
studied in Refs. \cite{white,allen,nersesyan,cabra,itoi}
and also in Ref. \cite{allenbis} in the S=1 case. 
In the special case of S=1/2 XY spins, it has been found by
Nersesyan et al.\cite{nersesyan} that
the model is a critical spin nematic.     
In this section, we shall investigate the existence
of such a phase in the general spin-S case and study its
stability as the interchain interaction is further varied.

\subsection{critical spin nematic phase}

The lattice Hamiltonian of 
the model (\ref{hamxyfrus}), considered as a two-leg spin ladder, 
is defined now as follows:
\bea
{\cal H} = \frac{J_2}{2} \sum_n
\left(S_{1,n}^{\dagger} S_{1,n+1}^{-} +
S_{2,n-1/2}^{\dagger} S_{2,n+1/2}^{-}
+ H.c. \right) \nonumber \\
+ \frac{J_1}{2} \sum_n
\left(S_{1,n}^{\dagger}\left(S_{2,n - 1/2}^{-}
+ S_{2,n + 1/2}^{-}\right)
+ H.c.
\right),
\label{zigzag1demi}
\eea
where $S_{1,n}^{\pm}$ (respectively $S_{2,n+1/2}^{\pm}$)
is the spin-S operator of chain of index $1$ (respectively $2$)
at site $n$ (respectively $n+1/2$).  
It is more suitable
to change the labeling of the second chain in the following way to
perform the continuum limit of the model: 
\bea
{\cal H} = \frac{J_2}{2} \sum_{a=1}^{2} \sum_n
\left(S_{a,n}^{\dagger} S_{a,n+1}^{-} + H.c. \right)
+ \frac{J_1}{2} \sum_n
\left(S_{1,n}^{\dagger}\left(S_{2,n}^{-} + S_{2,n-1}^{-}\right) + H.c.
\right).
\label{zigzag1demibis}
\eea
At this point, one should note that the interchain zigzag coupling
can also be written as (using intrachain periodic boundary conditions):
\be
{\cal H}_{int}^{'} = \frac{J_1}{2} \sum_n
\left(\left(S_{1,n}^{\dagger} + S_{1,n+1}^{\dagger}\right) S_{2,n}^{-} + H.c.
\right).
\ee
Consequently, we shall thus write the
interacting part of the Hamiltonian (\ref{zigzag1demibis})            
in a symmetrized way for taking the continuum limit of the model: 
\bea
{\cal H} =  \frac{J_2}{2} \sum_{a=1}^{2} \sum_n
\left(S_{a,n}^{\dagger} S_{a,n+1}^{-} + H.c. \right)  \nonumber \\
+ \frac{J_1}{4} \sum_n
\left(\left(S_{1,n}^{\dagger} + S_{1,n+1}^{\dagger}\right) S_{2,n}^{-} +
S_{1,n}^{\dagger}\left(S_{2,n}^{-} + S_{2, n-1}^{-}\right)
+ H.c. \right).
\label{zigzagxy1}
\eea
In the absence of the interchain coupling ($J_1 = 0$),
the model 
corresponds to two decoupled spin-S XY chains.
As seen in section II, 
it is critical with central charge $c=2$ and
its low-energy physics 
can be obtained with the introduction of two decoupled chiral gapless bosonic
fields $\Phi_{a+R,L}$ (a=1,2). 
The leading contribution of the spin density $S_a^{\pm}$
comes from the alternating part (see Eq. (\ref{spinSdensXY})):
\be
S_a^{\pm} \simeq \frac{\lambda}{\sqrt{a_0}} \left(-1\right)^{x/a_0}
\exp\left(\pm \ri \sqrt{\pi/2S}\; \;\Theta_{a+}\right),
\label{spinSdensXYbislad}
\ee                                
$\lambda$ being a non-universal constant.
From Eq. (\ref{spinSdensXYbislad}), we deduce the continuum limit of the 
model (\ref{zigzagxy1}) in the small $J_1 \ll J_2$ limit:
\bea
{\cal H} \simeq \frac{v}{2}\sum_{a=\pm}\left(\left(
\partial_x {\bar \Phi}_a\right)^2
+ \left(\partial_x {\bar \Theta}_a\right)^2\right)
+ g \; \partial_x 
{\bar \Theta}_+ \sin\left(\sqrt{\frac{\pi}{S}}{\bar \Theta}_-\right),
\label{zigzagcont1}
\eea                                              
where $g= J_1\lambda^2\sqrt{\pi/(4S)}$ and we have 
introduced the symmetric and antisymmetric combinations
of the two bosonic fields:
\bea 
{\bar \Phi}_{\pm} &=& \frac{1}{\sqrt{2}}\left(\Phi_{1+} \pm 
\Phi_{2+}\right) \nonumber \\ 
{\bar \Theta}_{\pm} &=& \frac{1}{\sqrt{2}}\left(\Theta_{1+} \pm 
\Theta_{2+}\right).
\label{basis}
\eea
The Hamiltonian (\ref{zigzagcont1}) describes a nontrivial field theory
since the field with coupling constant $g$, called 
twist term in Ref. \cite{nersesyan}, is a parity symmetry breaking
perturbation with a nonzero conformal spin (equal to one).
The effect of such term is rather unclear since the usual
irrelevant versus relevant criterion
does not hold for such a nonscalar perturbation
(see for instance Ref. \cite{book2}).
The simplest spin-1 conformal perturbation is the uniform part
of the spin density ($\partial_x \Phi$) that couples to a 
uniform magnetic field along the z-axis. In this case, 
this term leads to incommensuration as is well known.
It is thus natural to expect some incommensurability effect
in the model (\ref{zigzagcont1}) due to
the twist term as emphasized 
by Nersesyan et al.\cite{nersesyan}.                         
In particular, the presence of incommensuration
in the system can be found by a direct
mean-field analysis of the model (\ref{zigzagcont1}). 
Indeed, 
it is easy to see that the mean-field Hamiltonian separates
into two commuting parts:
${\cal H}_{MF} =  {\cal H}_+ + {\cal H}_-$ with
\bea
{\cal H}_+ &=& \frac{v}{2}\left(\left(\partial_x {\bar \Phi}_+\right)^2
+ \left(\partial_x {\bar \Theta}_+\right)^2\right) 
+ \kappa \partial_x {\bar \Theta}_+
\nonumber \\
{\cal H}_- &=& \frac{v}{2}\left(\left(\partial_x {\bar \Phi}_-\right)^2
+ \left(\partial_x {\bar \Theta}_-\right)^2\right) - \frac{\mu}{a_0}
\sin\left(\sqrt{\frac{\pi}{S}} {\bar \Theta}_-\right),
\label{meanf1}
\eea
the mean-field parameters being:
\bea
\kappa &=& g \langle \sin\left(\sqrt{\frac{\pi}{S}} 
{\bar \Theta}_-\right) \rangle
\nonumber \\
\frac{\mu}{a_0} &=& - g \langle \partial_x {\bar \Theta}_+ \rangle .
\label{meanfieldpara}
\eea                                                                       
The Hamiltonian (${\cal H}_+$) is easily solved by the redefinition
${\bar \Theta}_+ \rightarrow {\bar \Theta}_+ - \kappa x/v$.
The $+$ sector displays thus criticality with a nonzero topological
spin current in the ground state: 
$\langle 
\partial_x {\bar \Theta}_+ \rangle = - \kappa/v \ne 0$.
In contrast, the 
Hamiltonian (${\cal H}_-$) in the 
other sector is a standard sine Gordon 
model at $\beta^2 = \pi/S$ which describes a massive theory with 
massive quantum solitons and their bound states (breathers) together
with massive kinks. The dual field ${\bar \Theta}_-$ is locked at: 
$\langle {\bar \Theta}_-\rangle = \sqrt{\pi S/4} \; {\rm sgn} \mu \;
({\rm mod}\; \sqrt{4S\pi})$.
The mean-field analysis can then be closed using the fact that:
$\langle \sin(\sqrt{\pi/S}\; {\bar \Theta}_-) \rangle = 
c (a_0 |\mu|/v)^{1/(8S-1)}$ ($c$ being a constant that
can be determined\cite{luky}) and one easily finds: 
\bea 
\mu &=& \pm \frac{v}{a_0} \left(\frac{a_0 g}{v}\right)^{\frac{8S-1}{4S-1}} 
c^{\frac{8S-1}{8S-2}}
\nonumber \\
\kappa &=&  \pm \frac{v}{a_0} \left(\frac{a_0 g}{v}\right)^{\frac{4S}{4S-1}}
c^{\frac{8S-1}{8S-2}} .
\label{meanfieldferm}
\eea
From the correspondence (\ref{spinSdensXYbislad}), one can then estimate
the asymptotic behavior of the transverse spin-spin correlation functions
of the model which display an incommensurate critical behavior:
\be 
\langle S_1^{\dagger}\left(x\right) S^{-}_a\left(0\right) \rangle \sim 
\frac{\re^{iq_S x}}{|x|^{1/(8S)}}, \; \; a=1,2 ,
\label{correlzigzagS}
\ee
with $q_S-\pi/a_0 \sim (J_1/J_2)^{4S/(4S-1)}$.
The transverse spin-spin correlation functions fall off thus
with the distance as a power law with the exponent $1/(8S)$.
In the S=1 case,
one should note that
this exponent ($1/8 = 0.125$) 
found in this bosonization study is in good agreement with 
the numerical findings $0.15$ of the DMRG analyis of Ref. \cite{kawa1}.

Besides this incommensurate critical behavior observed
in the spin-spin correlation functions (\ref{correlzigzagS}), 
the physical picture of this phase obtained at the mean-field
level corresponds to a spin nematic\cite{andreev}.
Indeed, let us first introduce the z-component of the 
spin current $J_{as}^z$ associated to 
the ath spin-S XY chain ($a=1,2$):
\begin{equation}
J_{as}^{z} = -v \sqrt{\frac{2S}{\pi}} \; \partial_x \Theta_{a+}. 
\label{spincurrent}
\end{equation}
The vacuum expectation value of this operator can then be 
computed since one has in the ground state of 
the mean-field Hamiltonian (\ref{meanf1}):
$\langle
\partial_x {\bar \Theta}_+ \rangle = - \kappa/v \ne 0$ and 
$\langle
\partial_x {\bar \Theta}_- \rangle = 0$. This latter result
stems from the fact that the Hamiltonian (${\cal H}_-$) in 
Eq. (\ref{meanf1}) is a standard sine Gordon model characterized
by a ground state with zero topological charge. Using 
the redefinition (\ref{basis}), one finally obtains the following 
estimate:
\begin{equation}  
\langle J_{1s}^{z} \rangle = \langle J_{2s}^{z} \rangle 
= -v \sqrt{\frac{S}{\pi}} \; \langle \partial_x {\bar \Theta}_{+} \rangle
= \sqrt{\frac{S}{\pi}} \kappa \ne 0 .
\label{spincurrentave}
\end{equation}
These spin currents can also be expressed 
in terms of the original
spin degrees of freedom of the lattice Hamiltonian (\ref{zigzagxy1}) 
using the identification (\ref{spinSdensXYbislad}):
\begin{equation} 
\langle \left({\vec S}_{a,n} \wedge {\vec S}_{a,n+1}\right)_z \rangle \simeq 
- \lambda^2 \sqrt{\frac{\pi}{4S}}  \;
\langle \partial_x {\bar \Theta}_{+} \rangle \ne 0, \; \; a=1,2,
\label{chiralorderinchain}
\end{equation}
whereas similarily the (interchain) zigzag spin current along the z-axis
reads as follows:
\begin{equation} 
J_1 \langle \left({\vec S}_{1,n} \wedge {\vec S}_{2,n}\right)_z \rangle 
\simeq -2 \sqrt{\frac{S}{\pi}} g \langle \sin\left(\sqrt{\frac{\pi}{S}}
{\bar \Theta}_- \right) \rangle = - 2 \sqrt{\frac{S}{\pi}} \kappa \ne 0,
\label{chiralorderinterchain}
\end{equation}                 
where Eq. (\ref{meanfieldpara}) has been used.

The physical picture that emerges from this mean-field analysis
is therefore a spin nematic phase that preserves the U(1) and 
time reversal symmetries and displays long-range chiral ordering
in its ground state (\ref{chiralorderinchain}, \ref{chiralorderinterchain}).
In the classification of Ref. \cite{andreev}, this phase corresponds
to a p-type spin nematic.
At this point, it is important to stress that this chiral ordering
is different from the scalar chirality order operator\cite{wen}: 
$\langle 
{\vec S}_{1,n} . \left({\vec S}_{2,n} 
\wedge {\vec S}_{2,n-1} \right) \rangle$ which breaks parity and
time reversal symmetries. In our case, the spin nematic phase
does not break the time reversal symmetry but spontaneously breaks   
a $Z_2$ symmetry of the model which, as it will be shown in 
the next section, is a tensor product of a site-parity and link-parity
symmetries on the two chains. As a result, as first discovered
in the S=1/2 case in Ref. \cite{nersesyan}, this produces 
a picture of local nonzero 
spin currents (\ref{spincurrentave}, \ref{chiralorderinterchain}) 
polarized along
the z-anistropy axis circulating around the triangular plaquettes
of the two-leg zigzag spin ladder. 

\subsection{stability of the chiral critical phase}

It is important to study further
the stability of this critical spin nematic phase (chiral critical phase) 
in the $+$ channel
with respect to various
operators that will be generated in higher orders of
perturbation theory or equivalently terms consistent
with the symmetries of the original lattice model.      
Indeed, on general grounds, one expects that
some operators in the $+$ sector should destroy the criticality
of the phase at least for some finite value of $J_1/J_2$.
First of all, in the mean-field approach, the twist term
acts like a sort of magnetic field. As is well known,
a magnetic field along the anisotropy axis is a source of
incommensuration but also leads to a renormalization
of the compactification radius of the bosonic field.
This last effect was not found in the previous
approach as seen in the {\sl universal} behavior
of the spin-spin correlations (\ref{correlzigzagS}).
From a symmetry point of view (continuous U(1) symmetry),
there are no reasons to expect such universal behavior.
It is a first sign that higher order terms
in perturbation theory might be important here.
On the other hand, as seen in Section II, there is at least a massive
region (the dimerized or Haldane phases) 
in the phase diagram when increasing the
value of $J_1$ at fixed $J_2$. It is therefore likely that
a vertex operator, generated
in the renormalization group flow,
in the $+$ channel
will kill the critical phase at least for a critical value $(J_1/J_2)_c$.    

We shall now discuss the bosonic representation of the different 
discrete lattice symmetries of the model (\ref{zigzagxy1}) to 
find the nature of the operator that will be generated in the 
$+$ sector by the renormalization group flow. 
Let us first consider
the one-step translation ($t^{(a)}_{a_0}$),
site parity ($P_S^{(a)}$), and link parity  ($P_L^{(a)}$)
corresponding to the chain of index $a=1,2$. 
Using the definition (\ref{can}) of the diagonal bosonic field
that accounts for the criticality of the XY1 spin fluid phase 
in the 
decoupling limit ($J_1 =0$) 
and the bosonic 
representations (\ref{trans}, \ref{sparity}, \ref{lparity})
in the S=1/2 case described in the Appendix A,
one obtains the following identifications 
respectively for the one-step translation, site parity, and 
link parity:
\bea
\Phi_{a+} &\rightarrow& \Phi_{a+} 
+ \sqrt{\frac{\pi S}{2}} + p_a \sqrt{\frac{\pi}{2S}}\nonumber \\
\Theta_{a+} &\rightarrow& \Theta_{a+} + \sqrt{2S\pi} + p^{'}_a \sqrt{8\pi S},
\label{transs}
\eea
\bea
\Phi_{a+}\left(x\right) &\rightarrow& -\Phi_{a+}\left(-x\right)
+ \sqrt{\frac{\pi S}{2}} + q_a \sqrt{\frac{\pi}{2S}}\nonumber \\
\Theta_{a+}\left(x\right) &\rightarrow& \Theta_{a+}\left(-x\right)
+ q^{'}_a \sqrt{8\pi S},
\label{sparitys}
\eea
and
\bea
\Phi_{a+}\left(x\right) &\rightarrow& -\Phi_{a+}\left(-x\right) + 
n_a \sqrt{\frac{\pi}{2S}} 
\nonumber \\
\Theta_{a+}\left(x\right) &\rightarrow& \Theta_{a+}\left(-x\right)
+ \sqrt{2\pi S} +  n^{'}_a \sqrt{8\pi S},
\label{lparitys}
\eea
where $p_a, p^{'}_a, q_a, q^{'}_a, n_a, n^{'}_a$ are integers.
From these correspondences, one can deduce the 
bosonic representations of the discrete symmetries of 
the Hamiltonian (\ref{zigzagxy1}).
The translation symmetry acts on 
the symmetric and antisymmetric combinations (\ref{basis})  
of the bosonic fields $\Phi_{a+}$ as follows: 
\bea
{\bar \Phi}_+ &\rightarrow& {\bar \Phi}_+ + \sqrt{\pi S} +
\sqrt{\frac{\pi}{4S}}\left(p_1 + p_2\right) \nonumber \\
{\bar \Theta}_+ &\rightarrow& {\bar \Theta}_+ + \sqrt{4S\pi} +
\sqrt{4\pi S}\left(p_1^{'} + p_2^{'}\right) \nonumber \\
{\bar \Phi}_- &\rightarrow& {\bar \Phi}_- +
\sqrt{\frac{\pi}{4S}}\left(p_1 - p_2\right) \nonumber \\
{\bar \Theta}_- &\rightarrow& {\bar \Theta}_-  +
\sqrt{4\pi S}\left(p_1^{'} - p_2^{'}\right).
\label{transpinS}
\eea
A second type of discrete symmetry of the Hamiltonian (\ref{zigzagxy1}) $s_1$
consists of a vertical axial symmetry combined by an one-step translation
symmetry $t^{(1)}_{a_0}$ along the lower chain (labelled
$1$ in the following):
\bea
{\vec S}_{1,n} &\rightarrow& {\vec S}_{1,-n+1} \nonumber \\
{\vec S}_{2,n} &\rightarrow& {\vec S}_{2,-n},
\label{s1demi}
\eea
namely in the continuum limit:
\bea
{\vec n}_1\left(x\right) &\rightarrow& -{\vec n}_1\left(-x\right)\nonumber \\
{\vec n}_2\left(x\right) &\rightarrow& {\vec n}_2\left(-x\right),
\label{s1demicont}
\eea
which corresponds to a tensor product of a link-parity transformation
on chain $1$ and a site-parity transformation on chain $2$
($s_1 = P_L^{(1)} \otimes P_S^{(2)}$) when the model is viewed
as a zigzag ladder (Eq. (\ref{zigzag1demi})).                       
The bosonic representation of this 
discrete symmetry 
is thus:
\bea
{\bar \Phi}_+\left(x\right) &\rightarrow& -{\bar \Phi}_+\left(-x\right) 
+ \frac{\sqrt{\pi S}}{2} +
\sqrt{\frac{\pi}{4S}}\left(n_1 + q_2\right) \nonumber \\
{\bar \Theta}_+ \left(x\right)&\rightarrow& 
{\bar \Theta}_+\left(-x\right) + \sqrt{S\pi} +
\sqrt{4\pi S}\left(n_1^{'} + q_2^{'}\right) \nonumber \\
{\bar \Phi}_- \left(x\right) &\rightarrow& -{\bar \Phi}_-\left(-x\right) -
\frac{\sqrt{\pi S}}{2} +
\sqrt{\frac{\pi}{4S}}\left(n_1 - q_2\right) \nonumber \\
{\bar \Theta}_- \left(x\right) &\rightarrow& {\bar \Theta}_-\left(-x\right)
+ \sqrt{S\pi} + 
\sqrt{4\pi S}\left(n_1^{'} - q_2^{'}\right).                         
\label{s1S}
\eea                                   
In the same way, the Hamiltonian (\ref{zigzagxy1})
is also invariant under the transformation ($s_2$ symmetry):
\bea
{\vec S}_{1,n} &\rightarrow& {\vec S}_{1,-n}
\nonumber \\
{\vec S}_{2,n} &\rightarrow& {\vec S}_{2,-n-1},
\label{s2demi}
\eea
which can be viewed as a $P_L^{(2)} \otimes P_S^{(1)}$
transformation.                                
In terms of the bosonic fields
of the basis (\ref{basis}), this latter symmetry is
realized through:                 
\bea
{\bar \Phi}_+\left(x\right) &\rightarrow& -{\bar \Phi}_+\left(-x\right)
+ \frac{\sqrt{\pi S}}{2} +
\sqrt{\frac{\pi}{4S}}\left(n_2 + q_1\right) \nonumber \\
{\bar \Theta}_+ \left(x\right)&\rightarrow&
{\bar \Theta}_+\left(-x\right) + \sqrt{S\pi} +
\sqrt{4\pi S}\left(n_2^{'} + q_1^{'}\right) \nonumber \\
{\bar \Phi}_- \left(x\right) &\rightarrow& -{\bar \Phi}_-\left(-x\right) +
\frac{\sqrt{\pi S}}{2} +
\sqrt{\frac{\pi}{4S}}\left(-n_2 + q_1\right) \nonumber \\
{\bar \Theta}_- \left(x\right) &\rightarrow& {\bar \Theta}_-\left(-x\right)
- \sqrt{S\pi} +
\sqrt{4\pi S}\left(-n_2^{'} + q_1^{'}\right).
\label{s2S}
\eea                                       
There is a second family of discrete symmetries of
the Hamiltonian (\ref{zigzagxy1}):
$s_3 = P_{12} \otimes t^{(1)}_{a_0}$
or $s_4 =  P_{12} \otimes t^{(2)}_{-a_0}$
which corresponds to  an interchange of the chains combined
with a translation symmetry along the lower or upper chain.
In terms of the original spin degrees of freedom,
the $s_3$ and $s_4$ symmetries respectively write:
\bea
{\vec S}_{1,n} &\rightarrow& {\vec S}_{2,n}
\nonumber \\
{\vec S}_{2,n} &\rightarrow& {\vec S}_{1,n+1},
\label{s3demi}
\eea
\bea
{\vec S}_{1,n} &\rightarrow& {\vec S}_{2,n-1}
\nonumber \\
{\vec S}_{2,n} &\rightarrow& {\vec S}_{1,n},
\label{s4demi}
\eea
so that in the continuum limit, one has
\bea
{\vec n}_1\left(x\right) &\rightarrow& {\vec n}_2\left(x\right)\nonumber \\
{\vec n}_2\left(x\right) &\rightarrow& - {\vec n}_1\left(x\right),
\label{s3demicont}
\eea
and
\bea
{\vec n}_1\left(x\right) &\rightarrow& -{\vec n}_2\left(x\right)\nonumber \\
{\vec n}_2\left(x\right) &\rightarrow& {\vec n}_1\left(x\right) .
\label{s4demicont}
\eea                                                                 
The bosonic representation of these 
last discrete symmetries of Eq. (\ref{zigzagxy1}) 
is then respectively given by: 
\bea
{\bar \Phi}_+ &\rightarrow& {\bar \Phi}_+ + \frac{\sqrt{\pi S}}{2} +
\sqrt{\frac{\pi}{4S}} p_1 \nonumber \\
{\bar \Theta}_+ &\rightarrow& {\bar \Theta}_+ + \sqrt{S\pi} +
\sqrt{4\pi S} p_1^{'} \nonumber \\
{\bar \Phi}_- &\rightarrow& -{\bar \Phi}_- - \frac{\sqrt{\pi S}}{2}
- \sqrt{\frac{\pi}{4S}} p_1 \nonumber \\
{\bar \Theta}_- &\rightarrow& -{\bar \Theta}_-  - \sqrt{S\pi} -
\sqrt{4\pi S} p_1^{'},
\label{s3S}
\eea 
\bea
{\bar \Phi}_+ &\rightarrow& {\bar \Phi}_+ + \frac{\sqrt{\pi S}}{2} +
\sqrt{\frac{\pi}{4S}}   p_2 \nonumber \\
{\bar \Theta}_+ &\rightarrow& {\bar \Theta}_+ + \sqrt{S\pi} +
\sqrt{4\pi S} p_2^{'} \nonumber \\
{\bar \Phi}_- &\rightarrow& -{\bar \Phi}_- + \frac{\sqrt{\pi S}}{2}
+ \sqrt{\frac{\pi}{4S}} p_2 \nonumber \\
{\bar \Theta}_- &\rightarrow& -{\bar \Theta}_-  + \sqrt{S\pi} +
\sqrt{4\pi S} p_2^{'}.
\label{s4S}
\eea                              

With all these identifications, one
observes that the continuum limit (\ref{zigzagcont1})
of the lattice Hamiltonian (\ref{zigzagxy1})
is invariant under all discrete symmetries
(\ref{transpinS}, \ref{s1S}, \ref{s2S},
\ref{s3S}, \ref{s4S})
as it should be. However, the mean-field
Hamiltonian (\ref{meanf1}) is invariant under (\ref{transpinS},
\ref{s3S}, and \ref{s4S}) but
breaks the $s_{1,2}$ symmetries (\ref{s1S}, \ref{s2S}).  
These latter $Z_2$ discrete symmetries are spontaneously broken
in the ground state of the critical spin nematic phase and 
account for the formation of nonzero local spin currents
polarized along the z-axis circulating around the triangular
plaquette of the two-leg zigzag spin ladder.
The operator, that occurs in the $+$ sector
of the mean-field Hamiltonian (\ref{meanf1}),
with the smallest scaling dimension and 
consistent with
the symmetries (\ref{transpinS}, \ref{s3S}, \ref{s4S}) 
without breaking the continuous U(1) diagonal symmetry of the model
turns out to be:
$\cos(\mu \sqrt{16\pi S} {\bar \Phi}_+)$, with 
$\mu=1$ (respectively $\mu=2$) if S is integer (respectively
half-integer). The stable effective field theory in the $+$
channel is thus:
\bea
{\cal H}_+ \simeq
\frac{v}{2}\left(K \left(\partial_x {\bar \Theta}_+\right)^2
+ \frac{1}{K} \left(\partial_x {\bar \Phi}_+\right)^2 \right)
+ \kappa \partial_x {\bar \Theta}_+
- \frac{g_{eff}}{a_0} \cos\left(\mu \sqrt{16\pi S} {\bar \Phi}_+\right),
\label{effieldthSzigzag}
\eea                                          
where the value of the Luttinger parameters $v,K$ cannot
be determined within this bosonization approach.
For a small value of $J_1/J_2$ (i.e. $K \simeq 1$),
the cosine operator in Eq. (\ref{effieldthSzigzag})
is a strongly irrelevant contribution and the system displays a critical
phase with incommensuration generated by the $\partial_x {\bar \Theta}_+$ field.
This chiral critical phase, first predicted in the S=1/2 case
in  Ref. \cite{nersesyan}, is thus a generic phase in 
the large $J_1/J_2$ limit of the model (\ref{hamxyfrus})
in the general spin-S case. 
In particular, it is worth stressing that, in the S=1/2
case, the operator $\cos\left(\sqrt{8\pi} {\bar \Phi}_+\right)$,
which opens a mass gap in the $+$  channel
and thus destroys the chiral critical
phase found in Ref. \cite{nersesyan}, is not generated
by the renormalization group flow.
Indeed, while this latter operator is permitted by the 
translation symmetry (\ref{transpinS}),
it is odd under the $s_{3}$ and $s_{4}$ 
discrete symmetries
(\ref{s3S}, \ref{s4S})
which forbid its presence in the 
low-energy effective field theory.
This result leads us to expect that the chiral critical
phase does exist in the certain range of the parameter
of the lattice model for S=1/2 in full agreement with the very
recent numerical study\cite{nishi}.
As $J_1/J_2$ is further increased, 
it is  
natural to expect that the effective
theory (\ref{effieldthSzigzag}) describes
a phase transition of KT type from the chiral gapless phase at $g_{eff}=0$
to a chiral gapped phase. 
Indeed, there will be a critical value $(J_1/J_2)_c$
(the Luttinger parameter at the transition
being equal to $K_c = 1/(2S \mu^2)$), which
cannot be obtained within this bosonization approach,
above which the cosine operator $\cos(\beta {\bar \Phi}_+)$ 
becomes relevant 
and
a mass gap opens in the + sector
(KT transition) {\sl without} killing the incommensuration
stemming from the $\partial_x {\bar \Theta}_+$ operator.
In this respect, this mechanism of generation of incommensuration 
is different from the usual commensurate-incommensurate
scenario\cite{jnpt} since, in this latter case, there is 
a {\sl competition} between the uniform spin
density $\partial_x {\bar \Phi}_+$ field
and the cosine operator $\cos(\beta {\bar \Phi}_+)$ 
leading to a threshold above which
the incommensuration settles in the system.
One should note that
the existence of this incommensurate gapful phase 
when the cosine operator in Eq. (\ref{effieldthSzigzag})
becomes relevant can also been seen using
a Luther-Emery or Toulouse 
limit of the Hamiltonian (\ref{effieldthSzigzag})  
as it has been used to explain the origin of the incommensuration
found in the phase diagram of the quantum axial next-nearest-neighbor
Ising chain\cite{allen1}. 

The full characterization of the intermediate phase (chiral gapped
phase) depends on whether S is integer or half-integer.
Indeed, for $J_1/J_2 > (J_1/J_2)_c$,  the bosonic field 
${\bar \Phi}_+$ of Eq. (\ref{effieldthSzigzag}) is locked in 
one of the minima of the potential $- g_{eff} \cos(\mu \sqrt{16\pi S} 
{\bar \Phi}_+)$ which for $g_{eff} >0$\cite{commentbis} are located
at: $\langle {\bar \Phi}_+ \rangle = p\; \sqrt{\pi/4S}/\mu$, $p$
being an integer. Moreover, the value of the compactification radius
of the bosonic field ${\bar \Phi}_+$ is equal to: 
${\bar R}_S = 1/\sqrt{16\pi S}$. This follows from the
redefinition (\ref{basis}) and the fact that the compactification
radius of the bosonic field that accounts for the critical
properties of the spin-S XY chain is $R_S = 1/\sqrt{8\pi S}$
as it has been found in Section II.
From the precise knowledge of ${\bar R}_S$, one
deduces the following identification: 
\begin{equation} 
{\bar \Phi}_+ \sim {\bar \Phi}_+ + 2\pi {\bar R}_S = 
{\bar \Phi}_+ + \sqrt{\frac{\pi}{4S}}.
\end{equation}
From this equivalence and the 
position of the minima corresponding to 
the pinning of the bosonic field ${\bar \Phi}_+$, we thus 
conclude that in the integer spin case
($\mu =1$) the ground state of the massive phase is non-degenerate
whereas for half-integer spins ($\mu =2$) there is a two-fold
degenerate ground state\cite{commentiers}.
Therefore, the chiral gapful phase corresponds to 
a massive phase with a coexistence of incommensuration and 
a Haldane phase (respectively dimerized phase) in 
the integer (respectively half-integer) spin case.
From the identification of the massive phase
found at large $J_1/J_2$ in the weak coupling analysis (see
Section II), we then expect an Ising ($Z_2$) transition between
the chiral gapped phase and the Haldane or dimerized phases
as $J_1/J_2$ is further increased. At this Ising critical point,
the total spin current $\langle \partial_x {\bar \Theta}_+ \rangle$
vanishes i.e. the disappearance of the incommensurate behavior and the systems 
enters a commensurate massive phase: Haldane or dimerized phases
depending on the spin. At this point, one has to mention 
that the existence of this intermediate incommensurate massive
phase, within our mean-field approach, relies on 
the decouping of the degrees of freedom in 
the two channels $+$ and $-$ as in Eq. (\ref{meanf1}).
We cannot rule out a different scenario that might occur in the
system nonperturbatively due to the effect of the interactions
in the two sectors:
a single phase transition between the chiral critical phase
and the Haldane or dimerized phases.
At this critical point, one has {\sl simultaneously}
the appearance of a mass gap in the spectrum as well as 
the cancelation of the spin current so that the chiral 
gapful phase shrinks to zero in this case. 

\section{Concluding remarks}

In the present work, we have investigated 
the low-energy physics of the one-dimensional spin-S $J_1-J_2$ 
XY model within the bosonization approach.
Around the two limits ($J_2/J_1 \ll 1$, 
$J_1/J_2 \ll 1$) where a field theoretical analysis 
can be performed, 
we have described the nature of the different phases 
that occurs as well as the determination of
the effective field theories of the resulting 
phase transitions. 
The critical XY1 spin fluid phase at $J_2=0$
is generically stable upon switching on a nonzero value of the 
next-nearest-neighbor interaction except for the very
special S=1 case where the Haldane phase is immediately stabilized 
in full agreement with the DMRG study of Refs. \cite{kawa,kawa1}.
As the exchange interaction $J_2$ is further varied, the
model exhibits a KT phase transition described by 
a standard sine Gordon model between the XY1 spin fluid phase 
and a fully massive dimerized or Haldane phases 
depending on the value of the spin.
In the zigzag ladder limit ($J_1/J_2 \ll 1$),
we have shown that, whatever the value of the spin,
the chiral critical phase, first predicted in
the S=1/2 case by Nersesyan et al. \cite{nersesyan},
should exist in a certain range of the parameters of the
model.
This interesting spin nematic phase preserves the U(1)
and time-reversal symmetries but spontaneously 
breaks a $Z_2$ symmetry ($P_L^{(1)} \otimes P_S^{(2)}$)
resulting on the formation of nonzero local
spin currents in the ground state polarized along the anisotropy z-axis.  
Futhermore, the transverse spin-spin correlation functions are
incommensurate with a wave vector $q_S - \pi/a_0 
\sim (J_1/J_2)^{4S/(4S-1)}$
and decay algebraically with the distance with 
an exponent $1/(8S)$ obtained within the mean-field approach used here.
As the interchain $J_1/J_2$ is further increased, one expects 
the existence of a KT phase transition
between the chiral critical spin nematic phase and an 
incommensurate gapful phase (chiral gapped phase).
In particular,
the effective field theory corresponding to this transition 
has been determined in this work.
The nature of this chiral gapped phase corresponds to 
a coexistence of incommensuration stemming from the presence of nonzero
spin currents in the ground state and a Haldane or dimerized 
phases depending on whether the spin S is an integer or half-integer.
We then expect an Ising phase transition associated to the disappearance
of the spin current between the chiral gapped phase and the 
standard Haldane or dimerized phases.
The phase diagram found in this work is consistent with the predictions
of the large-S
study of Kolezhuk\cite{kol} except for the special S=1 case
where the XY1 spin fluid phase shrinks to zero.
It will be very interesting if some extended DMRG studies 
can be performed in the
$S>1$ case
to further shed light on the physical properties of the model
as well as the possibility to extract the Luttinger
parameters of the effective field theory (\ref{effieldthSzigzag}).
The different phase transitions reported in this work could
also be investigated by means of a level spectroscopy analysis
as in the one-dimensional spin-S XXZ Heisenberg model\cite{nomura2}. 

{\bf Acknowlegdments}

The authors would like to thank
D. Allen, E. Boulat, A. K. Kolezhuk, and A. A. Nersesyan 
for valuable discussions related to this work.                    

{\bf Note added} 

When this work was completed, we became aware of a very recent work
by Hikihara et al.\cite{hikihara} who have investigated 
the S=1/2,3/2,2 $J_1-J_2$ XY chain using a DMRG analysis.
They have found that the chiral critical phase 
appears in a broad region of the phase diagram in 
the general spin-S case in agreement with our work.
Futhermore, the prediction on the decay of the spin-spin
correlation ($1/(8S)$) in the chiral critical 
phase found within the bosonization approach
has been numerically verified.
Finally, for integer spins (S=1,2), the authors of Ref. \cite{hikihara} 
have reported the existence of the chiral gapped phase in a very
narrow region of the phase diagram whereas 
in the half-integer case (S=1/2,3/2) it 
has not been identified within the numerical precision of the
work \cite{hikihara}.

\appendix

\section*{A The XY chain in the continuum limit}             

In this Appendix, we shall recall some well known facts on the
continuum limit of the XY chain to fix the notations
that will be used throughout this paper.

The Hamiltonian of the antiferromagnetic
spin-1/2 XY chain is ($J_1 >0$):
\bea 
{\cal H}_0 =J_1  \sum_n\left(S^x_n S^x_{n+1} + S^y_n S^y_{n+1}\right),
\label{xy}
\eea
where ${\vec S}_n$ is a spin-1/2 operator at site n.
As is well known, this model can be written in terms 
of lattice fermions $c_n$ using the Jordan-Wigner transformation:
\bea
S_n^z &=& c_n^{\dagger} c_n - \frac{1}{2} \nonumber \\
S_n^{+} &=& (-1)^n c_n^{\dagger}
\exp\left(i\pi\sum_{j=1}^{n-1} c_j^{\dagger} c_j\right).
\label{jordanw}
\eea
The continuum limit of the model (\ref{xy}) can then be performed
with the introduction of 
right and left-moving fermion fields $R,L$:
$c_n/\sqrt{a_0} \rightarrow R(x) (i)^{x/a_0} + 
L(x) (-i)^{x/a_0}, x=na_0$, $a_0$ being the lattice spacing. 
Using the fermion-boson correspondence (see for instance 
Refs. \cite{book1,book2}): 
\bea 
R &=& \frac{1}{\sqrt{2\pi a_0}} 
\exp\left(i\sqrt{4\pi} \Phi_R\right) \nonumber \\
L &=& \frac{1}{\sqrt{2\pi a_0}} \exp\left(-i\sqrt{4\pi} \Phi_L\right),
\label{forbos}
\eea 
the Hamiltonian (\ref{xy}) can 
be expressed in terms of a bosonic field $\Phi$ and its
dual field $\Theta$ in the continuum limit:
\bea
{\cal H}_0 = \frac{v_0}{2}\int dx \left(\left(\partial_x \Theta\right)^2
+ \left(\partial_x \Phi\right)^2 \right),
\label{freeham}
\eea
where $v_0 = J_1 a_0$
is the spin velocity and 
we work with the following conventions:
\bea
\Phi = \Phi_L + \Phi_R \nonumber \\
\Theta = \Phi_L - \Phi_R \nonumber \\
\left[\Phi_R,\Phi_L\right] = i/4 .
\eea
This latter commutation relation is necessary to insure the 
anticommutation between the right and left fermion operators (see
Eq. (\ref{forbos})).
The bosonic field is compactified with the radius $R=1/\sqrt{4\pi}$: 
$\Phi \sim \Phi + \sqrt{\pi}$
whereas 
the dual field is compactified with the radius ${\tilde R} = 1/(2\pi R)$: 
$\Theta \sim \Theta + 2\sqrt{\pi}$.
The spin density operator in the continuum 
limit decomposes into uniform and alternating 
parts:
\be 
{\vec S} \simeq {\vec  J} + \left(-1\right)^{x/a_0} {\vec n},
\label{decompspin}
\ee
which 
can also be expressed in terms of the bosonic
fields as follows:
\bea 
n^z &=& \frac{-1}{\pi a_0}
\sin\left(\sqrt{4\pi}\Phi\right) \nonumber \\
n^{\dagger} &=& \frac{1}{\sqrt{2\pi a_0}}
\exp\left(\ri \sqrt{\pi}\Theta\right)  \nonumber \\ 
J^z &=& \frac{1}{\sqrt{\pi}} \partial_x \Phi \nonumber \\
J^{\dagger} &=& 
\frac{1}{\sqrt{8\pi a_0}}
\left(\exp\left(\ri 3 \sqrt{\pi} \Phi_L + \ri \sqrt{\pi} \Phi_R \right)
+\exp\left(-\ri 3 \sqrt{\pi} \Phi_R - \ri \sqrt{\pi} \Phi_L \right)
\right) \nonumber \\
&=& \frac{1}{\sqrt{2\pi a_0}} \exp\left(\ri \sqrt{\pi}\Theta\right)
\sin\left(\sqrt{4\pi}\Phi\right).
\label{contspinxy}
\eea

We end this appendix by giving the bosonic representation of 
the discrete symmetries of the XY Hamiltonian (\ref{xy})
that will be very useful when investigating
the stability of the chiral critical phase in Section III B.
Under a one-step translation symmetry,
the bosonic fields transform according to: 
\bea 
\Phi &\rightarrow& \Phi + \frac{\sqrt{\pi}}{2} + p \sqrt{\pi}\nonumber \\
\Theta &\rightarrow& \Theta + \sqrt{\pi} + p^{'} \sqrt{4\pi},
\label{trans}
\eea
$p, p^{'}$ being integers since from Eq. (\ref{decompspin})
the alternating part ($\vec n$) of the spin density 
should be odd under the one-step translation symmetry.
Under the site parity $P_s$ (${\vec S}_n\rightarrow {\vec S}_{-n}$),
the uniform and staggered parts of the spin density should be even so that: 
\bea
\Phi\left(x\right) &\rightarrow& -\Phi\left(-x\right) 
+ \frac{\sqrt{\pi}}{2} + q \sqrt{\pi}\nonumber \\
\Theta\left(x\right) &\rightarrow& \Theta\left(-x\right) 
+ q^{'} \sqrt{4 \pi}, 
\label{sparity}
\eea                
where $q, q^{'}$ are integers.
The link parity $P_L$ ($n\rightarrow 1-n$) 
is a combination of a site parity and a translation
symmetry so that under $P_L$ the bosonic fields $\Phi$ and $\Theta$ 
transform as:
\bea
\Phi\left(x\right) &\rightarrow& -\Phi\left(-x\right) + n \sqrt{\pi}
\nonumber \\ 
\Theta\left(x\right) &\rightarrow& \Theta\left(-x\right)
+ \sqrt{\pi} +  n^{'} \sqrt{4\pi}, 
\label{lparity}
\eea                                       
$n, n^{'}$ being integers.

\section*{B The one-dimensional S=1/2 $J_1-J_2$ XY model}

In this appendix, we derive 
the continuum limit of the model (\ref{hamxyfrus}) in
the S=1/2 case
in the weak coupling limit $J_2 \ll J_1$.
This calculation has been done several times\cite{haldane,nomura,zang} 
with different bosonized expressions.
This discrepancy stems from the fact 
that one has to be extremely careful when deriving
the continuum limit and in particular for 
obtaining the correct velocity renormalization. 
We shall redo here this calculation for  completeness
and also since it will be needed in Section II 
when deriving 
the bosonization approach of the $J_1-J_2$ spin-S XY chain 
in the $J_2 \ll J_1$ limit. 

The first step of the computation is to 
express the interacting part of Hamiltonian (\ref{hamxyfrus}) 
in terms of the lattice fermions
using the 
Jordan-Wigner transformation (\ref{jordanw}):
\be
{\cal H}_{int} = -J_2\sum_n\left(c_{n+2}^{\dagger}\left(
c_{n+1}^{\dagger} c_{n+1}-1/2\right) c_{n} + H.c.\right).
\ee
Using the continuum limit of the fermions and the bosonization
correspondence (\ref{forbos}) described in Appendix A, one has:
\bea 
{\cal H}_{int} = \frac{J_2 a_0}{2\pi} \int dx \left(
\left(-i\right)^{x/a_0} :\re^{-i\sqrt{4\pi} \Phi_R}:\left(x+2a_0\right)
\right. \nonumber \\
\left.
+ \left(i\right)^{x/a_0} :\re^{i\sqrt{4\pi} \Phi_L}:\left(x+2a_0\right)\right)
\nonumber \\
\left(\frac{1}{\sqrt{\pi}} \partial_x \Phi\left(x+a_0\right) 
+ \frac{\left(-1\right)^{x/a_0}}{\pi} :\sin\left(\sqrt{4\pi} \Phi\right):\left(x+a_0\right)\right)
\nonumber \\
\left(\left(i\right)^{x/a_0} :\re^{i\sqrt{4\pi} \Phi_R}:\left(x\right)
+ \left(-i\right)^{x/a_0} :\re^{-i\sqrt{4\pi} \Phi_L}:\left(x\right)\right) 
+ H. c.
\label{contweak1demi}
\eea
To derive the continuum expression of this Hamiltonian, 
we need the following operator product
expansions in a standard gaussian $c=1$ theory:
\bea 
:\re^{-i\sqrt{4\pi} \Phi_R}:\left(\bar z\right) 
\partial_x \Phi\left(w,\bar w\right) \sim \;
:\partial_x \Phi_L \re^{-i\sqrt{4\pi} \Phi_R}:\left(w,\bar w\right) 
\nonumber \\
-\frac{1}{\sqrt{4\pi}\left(\bar z - \bar w\right)}
:\re^{-i\sqrt{4\pi} \Phi_R}: \left(\bar w\right) 
- \frac{1}{\sqrt{4\pi}}
:{\bar \partial}\re^{-i\sqrt{4\pi} \Phi_R}: \left(\bar w\right) \nonumber \\ 
\nonumber \\
:\re^{i\sqrt{4\pi} \Phi_L}:\left(z\right)
\partial_x \Phi\left(w,\bar w\right) \sim \;
:\partial_x \Phi_R \re^{i\sqrt{4\pi} \Phi_L}:\left(w,\bar w\right)
\nonumber \\
-\frac{1}{\sqrt{4\pi}\left(z - w\right)}
:\re^{i\sqrt{4\pi} \Phi_L}: \left(w\right)
- \frac{1}{\sqrt{4\pi}}
:\partial\re^{i\sqrt{4\pi} \Phi_L}: \left(w\right)  \nonumber \\
\nonumber \\
:\re^{-i\sqrt{4\pi} \Phi_R}:\left(\bar z\right)     
:\sin\left(\sqrt{4\pi} \Phi\right):\left(w,\bar w\right) \sim \;
-\frac{\bar z - \bar w}{2} :\re^{-i\sqrt{16\pi} \Phi_R\left(\bar w\right)}
\re^{-i\sqrt{4\pi} \Phi_L\left(w\right)}: \nonumber \\
+ \frac{1}{2\left(\bar z- \bar w\right)} :\left(1 - i\sqrt{4\pi} \left(\bar z - 
\bar w\right) {\bar \partial} \Phi_R - 2\pi\left(\bar z - \bar w\right)^2
\left({\bar \partial} \Phi_R\right)^2\right) \re^{i\sqrt{4\pi} \Phi_L}:
\left(w,\bar w\right) \nonumber \\
\nonumber \\
:\re^{i\sqrt{4\pi} \Phi_L}:\left(z\right)
:\sin\left(\sqrt{4\pi} \Phi\right):\left(w,\bar w\right) \sim \;
-\frac{z - w}{2} :\re^{i\sqrt{16\pi} \Phi_L\left(w\right)}
\re^{i\sqrt{4\pi} \Phi_R\left(\bar w\right)}: \nonumber \\
+ \frac{1}{2\left(z- w\right)} :\left(1 + i\sqrt{4\pi} \left(z -
w\right) \partial \Phi_L - 2\pi\left(z - w\right)^2
\left(\partial \Phi_L\right)^2\right) \re^{-i\sqrt{4\pi} \Phi_R}:
\left(w,\bar w\right)                                             
\label{someope}
\eea
with the convention $w= v_0 \tau + i x$ and $\partial_x = i (\partial -
{\bar \partial})$.
Using these results and keeping only 
non-oscillatory contributions in Eq. (\ref{contweak1demi}), 
we finally obtain: 
\bea 
{\cal H}_{int} \simeq -\frac{J_2}{\pi^2 a_0} \int d x 
\cos\left(\sqrt{16\pi} \Phi\right) - \frac{4J_2 a_0}{\pi} 
\int d x \left(\partial_x \Theta\right)^2
\label{contxy2}
\eea
which is in perfect agreement
with the earlier derivation made by Haldane (see the erratum)\cite{haldane}
and is in contradiction with some recent ones 
in the litterature\cite{nomura,zang}.


\begin{thebibliography}{99}
\bibitem{comment} When the cosine term $\cos(\mu \sqrt{8\pi S} \Phi_+)$
in Eq. (\ref{effieldthS}) becomes relevant, the bosonic 
field $\Phi_+$ becomes pinned in one of the minima of the 
sine Gordon model at $\beta^2 = \mu \sqrt{8\pi S}$.
The effective coupling constant $g_{eff}$ that enters
in Eq. (\ref{effieldthS}) cannot be fixed within our approach.
In some situations, the sign of this coupling
constant is crucial to fully determine the 
nature of the discrete symmetry breaking that may occur
in a massive phase but here it will not be important.  
The position of the minima of the 
interacting part of Eq. (\ref{effieldthS}) for $g_{eff} > 0$ 
reads as follows: $\langle \Phi_+ \rangle = p \sqrt{\pi/(2S)}/\mu$, $p$ being an
integer. However, the bosonic field $\Phi_+$ has a compactification
radius equals to $R_S = 1/\sqrt{8\pi S}$ so that one has
the following equivalence: $\Phi_+ \sim \Phi_+ + \sqrt{\pi/(2S)}$.
From the position of the minima where the $\Phi_+$ field is locked,
we thus deduce that in 
the integer spin case ($\mu =1$) the ground state in the
massive phase is non-degenerate whereas in the half-integer
case ($\mu =2$) there is a two-fold degenerate ground state
as it should be respectively for the Haldane and dimerized phases. 
In the latter case, there is a $Z_2$ symmetry breaking 
corresponding to the symmetry breaking of one-site lattice 
translation. Finally, if one has $g_{eff} < 0$, it is easy to see 
that the same physical characterization of the 
massive phase can be obtained.
\bibitem{commentbis} The sign of this coupling constant cannot
be fixed within our approach since the cosine term that
occurs in Eq. (\ref{effieldthSzigzag}) was obtained through 
a symmetry analysis.
However, the precise sign of $g_{eff}$ is irrelevant
for the determination of the nature of the incommensurate
gapful phase.
\bibitem{commentiers} Here we do not take into account of the 
additional two-fold degeneracy
of the ground state that stems from the spontaneous breaking
of a $Z_2$ discrete symmetry (\ref{s1demi}) of the lattice model
associated to the formation of 
nonzero spin currents in the system.
\end{thebibliography}
\end{document}